\documentclass[12pt]{article}
\usepackage{amsmath,amsfonts,amssymb}

\begin{document}
\thispagestyle{empty}

\def\theequation{\arabic{section}.\arabic{equation}}
\def\a{\alpha}
\def\b{\beta}
\def\g{\gamma}
\def\d{\delta}
\def\dd{\rm d}
\def\e{\epsilon}
\def\ve{\varepsilon}
\def\z{\zeta}
\def\B{\mbox{\bf B}}

\newcommand{\h}{\hspace{0.5cm}}

\begin{titlepage}
\vspace*{1.cm}
\renewcommand{\thefootnote}{\fnsymbol{footnote}}
\begin{center}
{\Large \bf Semiclassical structure constants in the
$\eta$-deformed $AdS_5\times S^5$: Leading finite-size
corrections}
\end{center}
\vskip 1.2cm \centerline{\bf Plamen Bozhilov} \vskip 0.6cm
\centerline{\sl Institute for Nuclear Research and Nuclear Energy}
\centerline{\sl Bulgarian Academy of Sciences} \centerline{\sl
1784 Sofia, Bulgaria}

\centerline{\tt bozhilov@inrne.bas.bg, bozhilov.p@gmail.com}

\vskip 20mm

\baselineskip 18pt

\begin{center}
{\bf Abstract}
\end{center}

\h We consider the leading finite-size effects on some structure
constants for the $\eta$ - deformed $AdS_5\times S^5$ background
in the framework of the semiclassical approach. The leading
finite-size corrections are derived for the cases when we have two
heavy string states represented by giant magnons and for two
different choices of the light states corresponding to dilaton
operator with nonzero momentum and primary scalar operators. Since
the dual field theory is still unknown, the results obtained here
must be considered as conjectures or as predictions from the
string theory side.

\end{titlepage}
\newpage

\def\nn{\nonumber}
\def\tr{{\rm tr}\,}
\def\p{\partial}
\newcommand{\bea}{\begin{eqnarray}}
\newcommand{\eea}{\end{eqnarray}}
\newcommand{\bde}{{\bf e}}
\renewcommand{\thefootnote}{\fnsymbol{footnote}}
\newcommand{\be}{\begin{equation}}
\newcommand{\ee}{\end{equation}}

\vskip 0cm

\renewcommand{\thefootnote}{\arabic{footnote}}
\setcounter{footnote}{0}


\setcounter{equation}{0}
\section{Introduction}
\h In the framework of the AdS/CFT duality \cite{AdS/CFT} between
string theories/M-theory on curved space-times with Anti-de Sitter
subspaces and conformal field theories in different dimensions, an
interesting issue to work on is to go beyond the spectral
problem\footnote{For recent review see \cite{RO}.} and compute the
corresponding correlation functions, if possible.

It was shown in \cite{J1002}  how to compute a two-point
correlation function of operators, dual to classical spinning
strings in a subspace of $AdS_5\times S^5$. These investigations
have been continued in \cite{BT2010}, where the 2-point function
of string vertex operators representing string state with large
spin in $AdS_5$ have been obtained in the semiclassical
approximation.

The computation of semiclassical three-point correlation functions
in $AdS_5\times S^5$ string theory, dual to $\mathcal{N}=4$ SYM
in four space-time dimensions, was started in
\cite{Z1008}-\cite{rt10} followed by many other papers.

A new development was to consider deformations of $AdS_5\times
S^5$ and obtain some three-point correlators (structure
constants). This was done first for the $\gamma$-deformed
\cite{LM05} or $TsT$-transformed \cite{F05} $AdS_5\times S^5$ dual
to $\mathcal{N}=1$ SYM. The first papers on this subject were
\cite{AR1106,AB1106}.

Recently, a new {\it integrable} deformation ($\eta$-deformation)
of the $AdS_5 \times S^5$ superstring action has been discovered
\cite{DMV0913}\footnote{$\eta$-deformed $\sigma$-models where
defined and their integrability was proven in \cite{K02,K08}. See
also \cite{K15}.}.
The bosonic part of the superstring sigma model Lagrangian on this
$\eta$-deformed background was determined in \cite{ABF1312}. From
it one can extract the background metric $g$ and the 2-form gauge
potential $b$ \cite{AP2014}. Then, one can find different string
solutions on this background, e.g. the ones found in
\cite{AP2014}-\cite{BBP15} (for other contributions on the subject
see the references in \cite{B0215}). In particular, giant magnon
type solution, playing an essential role in the string
theory-field theory duality, has been found \cite{ALT1403}. Based
on this solution, one can compute the corresponding vertices on
it. Then, a natural task is to try to find the three-point
correlation functions (structure constants) for two heavy string
states represented by giant magnons and some light (supergravity)
states. The first step in this direction was done in
\cite{AB1412}, where the finite-size effect was also taken into
account for the case of two giant magnon states and dilaton
operator with zero momentum. This result was extended in
\cite{B0215} for the cases of two giant magnon states and dilaton
operator with nonzero momentum, primary scalar operators, and
singlet scalar operators on higher string levels.

The semiclassical structure constants found in \cite{B0215} are
given in terms of several parameters and hypergeometric functions
of two variables depending on them. However, it is important to
know their dependence on the conserved string charge $J_1$ and the
worldsheet momentum $p$, because namely these quantities are
related to the corresponding operators in the dual field theory
and the momentum of the magnon excitations in the dual spin-chain.
Unfortunately, this can not be done exactly for the finite-size
case due to the complicated dependence between the above mentioned
parameters and $J_1$, $p$. On the other hand, it is possible to
find the leading finite-size corrections to such structure
constants. Namely this will be our aim here.

This letter is organized as follows. In Sec. 2 we give a short
review of the finite-size giant magnon solution on $\eta$-deformed
$AdS_5\times S^5$. In Sec. 3 and Sec. 4 we derive the leading
finite-size effects on the structure constants for the cases when
we have two heavy string states represented by finite-size giant
magnons and for two different choices of the light states
corresponding to dilaton operator with nonzero momentum and
primary scalar operators. Sec. 5 is devoted to our concluding
remarks.

\setcounter{equation}{0}
\section{Short review of the giant magnon solution \\ on $\eta$-deformed $AdS_5\times S^5$}

Giant magnons live in the $R_t\times S^2_{\eta}$ subspace of the
$\eta$-deformed $AdS_5\times S^5$. The background seen by the
string moving in the $R_t\times S^2_\eta$ subspace can be written
as \cite{AP2014} \bea\nn &&g_{tt}=-1,\h
g_{\phi_1\phi_1}=\sin^2\theta,
\\ \label{fb} &&g_{\theta\theta}=\frac{1}{1+\tilde{\eta}^2 \sin^2\theta},\eea
 where
$\tilde{\eta}$ is related to the deformation parameter $\eta$
according to\footnote{We changed the notation $\kappa$ in
\cite{ALT1403} to $\tilde{\eta}$ because we use $\kappa$ for other
purposes.} \bea\label{ek} \tilde{\eta}=\frac{2
\eta}{1-\eta^2}.\eea

By using a specific anzatz for the string embedding \cite{KRT06}
\bea\nn &&t(\tau,\sigma)=\kappa\tau,\h
\phi_1(\tau,\sigma)=\omega_1\tau+F(\xi),\h
\theta(\tau,\sigma)=\theta(\xi),\h \xi=\alpha\sigma+\beta\tau,
\\ \nn &&\kappa, \omega_1, \alpha, \beta=constants,\eea
one can find the following string solution
\cite{AP2014}\footnote{See \cite{AR1406} where it was shown that
the bosonic spinning strings on the $\eta$-deformed $AdS_5 x S^5$
background are naturally described as periodic solutions of a
novel finite-dimensional integrable system which can be viewed as
a deformation of the Neumann model.} \bea\label{chis} \chi(\xi)=
\frac{\chi_\eta\chi_p\ \mathbf{dn}^2(x,m)}{\chi_p \
\mathbf{dn}^2(x,m)+\chi_\eta-\chi_p},\eea

\bea\label{f1s} &&\phi_1(\tau,\sigma)= \omega_1
\tau+\frac{1}{\tilde{\eta}\alpha\omega_1^2(\chi_\eta-1)
\sqrt{(\chi_\eta-\chi_m)\chi_p}} \times
\\ \nn &&\Bigg\{\Big[\beta\left(\kappa^2+\omega_1^2(\chi_\eta-1)\right)
\Big]\
\mathbf{F}\left(\arcsin\sqrt{\frac{(\chi_\eta-\chi_m)(\chi_p-\chi)}{(\chi_p-\chi_m)(\chi_\eta-\chi)}},
m\right)
\\ \nn &&-\frac{(\chi_\eta-\chi_p)\beta\kappa^2}{1-\chi_p}\
\mathbf{\Pi}\left(\arcsin\sqrt{\frac{(\chi_\eta-\chi_m)(\chi_p-\chi)}{(\chi_p-\chi_m)(\chi_\eta-\chi)}}
,-\frac{(\chi_\eta-1)(\chi_p-\chi_m)}{(1-\chi_p)(\chi_\eta-\chi_m)},m\right)\Bigg\}.\eea

Here $\chi=\cos^2\theta$, where $\theta$ is the non-isometric
angle on the deformed sphere $S^2_{\eta}$, while $\phi_1$ is the
isometric angle on it. $\mathbf{dn}(x,m)$ is one of the Jacobi
elliptic functions, $\mathbf{F}$ and $\mathbf{\Pi}$ are the
incomplete elliptic integrals of first and third kind. \bea\nn
&&x=\frac{\tilde{\eta} \alpha
\omega_1\sqrt{(\chi_\eta-\chi_m)\chi_p}} {\alpha^2-\beta^2}\ \xi ,
\\ \nn &&m= \frac{(\chi_p-\chi_m)\chi_\eta}{(\chi_\eta-\chi_m)\chi_p},\eea
and $\chi_\eta >\chi_p>\chi_m$ are the roots of the equation $d
\chi/d \xi=0$, given by \bea\label{roots}
\chi_\eta=1+\frac{1}{\tilde{\eta}^{2}}\h
\chi_p=1-\frac{\beta^2}{\alpha^2} \kappa^2,\h
\chi_m=1-\kappa^2.\eea

Now, let us present the expressions for the conserved charges (the
string energy $E_s$ and the angular momentum $J_1$) and also the
worldsheet momentum $p$ equal to the angular difference
$\Delta\phi_1$, since we are going to use it \cite{AP2014}:
\bea\label{Esi}
E_s=\frac{T}{\tilde{\eta}}\left(1-\frac{\beta^2}{\alpha^2}\right)
\frac{\kappa}{\omega_1} \int_{\chi_m}^{\chi_p}
\frac{d\chi}{\sqrt{(\chi_\eta-\chi)(\chi_p-\chi)(\chi-\chi_m)\chi}},\eea
\bea\label{J1i} J_1&=&\frac{T}{\tilde{\eta}}
\left[\left(1-\frac{\beta^2\kappa^2}{\alpha^2\omega_1^2}\right)
\int_{\chi_m}^{\chi_p}
\frac{d\chi}{\sqrt{(\chi_\eta-\chi)(\chi_p-\chi)(\chi-\chi_m)\chi}}\right.
\\ \nn &&-\left.\int_{\chi_m}^{\chi_p}
\frac{\chi
d\chi}{\sqrt{(\chi_\eta-\chi)(\chi_p-\chi)(\chi-\chi_m)\chi}}\right],\eea

\bea\label{adi} \Delta\phi_1\equiv p &=&\frac{1}{\tilde{\eta}}
\Bigg[\frac{\beta}{\alpha}\int_{\chi_m}^{\chi_p}
\frac{d\chi}{\sqrt{(\chi_\eta-\chi)(\chi_p-\chi)(\chi-\chi_m)\chi}}
\\ \nn &&-\frac{\beta\kappa^2}{\alpha\omega_1^2}\int_{\chi_m}^{\chi_p}
\frac{d\chi}{(1-\chi)\sqrt{(\chi_\eta-\chi)(\chi_p-\chi)(\chi-\chi_m)\chi}}\Bigg].\eea
Solving the integrals in (\ref{Esi})-(\ref{adi}) and introducing
the notations \bea\label{not} v=-\frac{\beta}{\alpha},\
W=\frac{\kappa^2}{\omega_1^2},\
\epsilon=\frac{(\chi_\eta-\chi_p)\chi_m}{(\chi_\eta-\chi_m)\chi_p},\eea
we finally obtain \bea\label{Esf} E_s=\frac{2T}{\tilde{\eta}}
\frac{(1-v^2)\sqrt{W}}{\sqrt{(\chi_\eta-\chi_m)\chi_p}} \
\mathbf{K}(1-\epsilon),\eea \bea\label{J1f}
J_1&=&\frac{2T}{\tilde{\eta}\sqrt{(\chi_\eta-\chi_m)\chi_p}}
\Bigg[\left(1-v^2W-\chi_\eta\right)\ \mathbf{K}(1-\epsilon)
\\ \nn &&+(\chi_\eta-\chi_p)\ \mathbf{\Pi}\left(\frac{\chi_p-\chi_m}{\chi_\eta-\chi_m},
1-\epsilon\right)\Bigg],\eea

\bea\label{adf} \Delta\phi_1\equiv p &=&
\frac{2}{\tilde{\eta}\sqrt{(\chi_\eta-\chi_m)\chi_p}}\times
\\ \nn &&\Bigg\{\frac{v W}{(\chi_\eta-1)(1-\chi_p)}\Bigg[(\chi_\eta-\chi_p)\
\mathbf{\Pi}\left(-\frac{(\chi_\eta-1)(\chi_p-\chi_m)}{(\chi_\eta-\chi_m)(1-\chi_p)},
1-\epsilon\right)
\\ \nn &&-(1-\chi_p)\ \mathbf{K}(1-\epsilon)\Bigg]-v \ \mathbf{K}(1-\epsilon)\Bigg\},\eea
where $\mathbf{K}$ and $\mathbf{\Pi}$ are the complete elliptic
integrals of first and third kind.

The above results are for finite-size giant magnons. If they are
of infinite size, one must set $\epsilon=0$.

The spectrum of a string moving on $\eta$-deformed $AdS_5 \times
S^5$ was considered in \cite{ALT1403}. This was done by treating
the corresponding worldsheet theory as integrable field theory. In
particular, it was found that the dispersion relation for the
infinite-size giant magnons \cite{HM06} on this background, in the
large string tension limit when $g \to \infty$ is given by (the
relation between the string tension $T$ and $g$ is 
$T=g\sqrt{1+\tilde{\eta}^2}$)

\bea\label{14} E=\frac{2g\sqrt{1+\tilde{\eta}^2}}{\tilde{\eta}}
\mbox{arcsinh}\left(\tilde{\eta} \sin\frac{p}{2}\right).\eea

The result (\ref{14}) has been extended in \cite{AP2014} to the
case of finite-size giant magnons. The corresponding dispersion
relation is the following \bea\label{fr} E_s-J_1= 2 g
\sqrt{1+\tilde{\eta}^2} \left[\frac{1}{\tilde{\eta}}
\mbox{arcsinh}\left(\tilde{\eta}
\sin\frac{p}{2}\right)-\frac{(1+\tilde{\eta}^2)
\sin^3\frac{p}{2}}{4 \sqrt{1+\tilde{\eta}^2 \sin^2\frac{p}{2}}}\
\epsilon\right],\eea where
\bea\label{eps} \epsilon =16\
\exp\left[-\left(\frac{J_1}{g}
+\frac{2\sqrt{1+\tilde{\eta}^2}}{\tilde{\eta}}\mbox{arcsinh}\left(\tilde{\eta}
\sin\frac{p}{2}\right)
\right)\sqrt{\frac{1+\tilde{\eta}^2\sin^2\frac{p}{2}}{\left(1+\tilde{\eta}^2\right)\sin^2\frac{p}{2}}}
\right].\eea

\setcounter{equation}{0}
\section{Giant magnons on $\eta$-deformed $AdS_5\times S^5$ and dilaton operator}
The case of finite-size giant magnons and dilaton with zero
momentum $(j=0)$ has been considered in \cite{AB1412}. Here we
will be interested in the case when $j>0$.

In \cite{B0215} it was found that the normalized semiclassical
structure constants for the case under consideration are given by
\bea\label{etadj} \mathcal{C}_{\eta}^{d,j}&=&\frac{2
\pi^{\frac{3}{2}}c_\Delta^{d,j}
\Gamma\left(2+\frac{j}{2}\right)(1-v^2 \kappa^2)^{\frac{j-1}{2}}}
{\Gamma\left(\frac{5+j}{2}\right)\sqrt{\kappa^2(1+\tilde{\eta}^2
\kappa^2)}}\times
\\ \nn
&&\Bigg[(1-v^2 \kappa^2)F_1
\left(\frac{1}{2},\frac{2+j}{2},-\frac{1+j}{2};1;\frac{\tilde{\eta}^2(1-v^2)\kappa^2}
{1+\tilde{\eta}^2\kappa^2},\frac{(1+\tilde{\eta}^2)(1-v^2)\kappa^2}
{(1+\tilde{\eta}^2\kappa^2)(1-v^2\kappa^2)}\right)
\\ \nn &&-(1-\kappa^2)F_1 \left(\frac{1}{2},\frac{j}{2},\frac{1-j}{2};1;\frac{\tilde{\eta}^2(1-v^2)\kappa^2}
{1+\tilde{\eta}^2\kappa^2},\frac{(1+\tilde{\eta}^2)(1-v^2)\kappa^2}
{(1+\tilde{\eta}^2\kappa^2)(1-v^2\kappa^2)}\right)\Bigg],\eea
where $F_1(a,b_1,b_2;c;z_1,z_2)$ is one of the hypergeometric
functions of two variables ($AppellF_1$).

This expression can be rewritten as (according to (\ref{not})
$W=\kappa^2$ for $\omega_1=1$)

\bea\label{etadjW} \mathcal{C}_{\eta}^{d,j}&=&\frac{2
\pi^{\frac{3}{2}}c_\Delta^{d,j}
\Gamma\left(2+\frac{j}{2}\right)(1-v^2 W)^{\frac{j-1}{2}}}
{\Gamma\left(\frac{5+j}{2}\right)\sqrt{W(1+\tilde{\eta}^2
W)}}\times
\\ \nn
&&\Bigg[(1-v^2W)F_1
\left(\frac{1}{2},\frac{2+j}{2},-\frac{1+j}{2};1;\frac{\tilde{\eta}^2(1-v^2)W}
{1+\tilde{\eta}^2W},1-\epsilon\right)
\\ \nn &&-(1-W)F_1 \left(\frac{1}{2},\frac{j}{2},\frac{1-j}{2};1;\frac{\tilde{\eta}^2(1-v^2)W}
{1+\tilde{\eta}^2W},1-\epsilon\right)\Bigg],\eea

\bea\label{epsW}
\epsilon=\frac{(1-W)(1+\tilde{\eta}^2v^2W)}{(1-v^2W)(1+\tilde{\eta}^2W)}
.\eea

In order to find the leading finite-size effect on (\ref{etadjW})
one has to consider the limit $\epsilon\to 0$. By using the
following expansions for the parameters \bea\label{vW}
v=v_0+(v_1+v_2\log\epsilon)\epsilon,\h W=1+W_1\epsilon,\eea one
obtains \cite{AP2014} (by using (\ref{roots}) and (\ref{adf}))

\bea\label{W1}
W_1=-\frac{(1-v_0^2)(1+\tilde{\eta}^2)}{1+\tilde{\eta}^2v_0^2},\eea

\bea\label{v1} v_1=\frac{v_0(1-v_0^2)\left[1-\log
16+\tilde{\eta}^2\left(2-v_0^2(1+\log
16)\right)\right]}{4(1+\tilde{\eta}^2v_0^2)},\eea

\bea\label{v2} v_2=\frac{1}{4}v_0(1-v_0^2),\eea

where $v_0$ can be written as \bea\label{v0}
v_0&=&\frac{\cot\frac{\Delta\phi_1}{2}}{\sqrt{\tilde{\eta}^2+\csc^2\frac{\Delta\phi_1}{2}}}
=\frac{\cot\frac{p}{2}}{\sqrt{\tilde{\eta}^2+\csc^2\frac{p}{2}}}.\eea

Applying these results for the parameters in the
$\epsilon$-expansion of (\ref{etadjW}), one receives the following
expression \bea\label{fcd}  &&\mathcal{C}_{\eta}^{d,j}\approx
\frac{ c_\Delta^{d,j}\pi
\Gamma\left(2+\frac{j}{2}\right)\Gamma\left(\frac{j}{2}\right)}
{\Gamma\left(\frac{5+j}{2}\right)\Gamma\left(\frac{1+j}{2}\right)}
(1+\tilde{\eta}^2)^{j/2}\left(\tilde{\eta}^2+\csc^2\frac{p}{2}\right)^{-\frac{1}{2}(1+j)}
\\ \nn &&\Bigg[\frac{(2+\tilde{\eta}^2(1-\cos p))\csc^2\frac{p}{2}}
{\tilde{\eta}^2}\Bigg({}_2F_1\left(\frac{1}{2},\frac{j}{2};\frac{1+j}{2};
\frac{1}{1+\frac{\csc^2\frac{p}{2}}{\tilde{\eta}^2}}\right)
\\ \nn &&-{}_2F_1\left(-\frac{1}{2},\frac{j}{2};\frac{1+j}{2};
\frac{1}{1+\frac{\csc^2\frac{p}{2}}{\tilde{\eta}^2}}\right)\Bigg)
\\ \nn &&+4\Bigg(\Bigg(\frac{1}{\tilde{\eta}^2}\Bigg(2(2+\tilde{\eta}^2(1-\cos p))
{}_2F_1\left(\frac{1}{2},\frac{j}{2};\frac{1+j}{2};
\frac{1}{1+\frac{\csc^2\frac{p}{2}}{\tilde{\eta}^2}}\right)
\\ \nn &&-(4+\tilde{\eta}^2(6-j+2\tilde{\eta}^2-(2+j+2\tilde{\eta}^2)\cos p))
{}_2F_1\left(-\frac{1}{2},\frac{j}{2};\frac{1+j}{2};
\frac{1}{1+\frac{\csc^2\frac{p}{2}}{\tilde{\eta}^2}}\right)
\Bigg)\Bigg) \\ \nn &&+\Bigg(j(1+\cos
p){}_2F_1\left(-\frac{1}{2},\frac{j}{2};\frac{1+j}{2};
\frac{1}{1+\frac{\csc^2\frac{p}{2}}{\tilde{\eta}^2}}\right)\Bigg)f\Bigg)\exp(-f)\Bigg],\eea
where \bea\label{f} f=\left(\frac{J_1}{g}
+\frac{2\sqrt{1+\tilde{\eta}^2}}{\tilde{\eta}}\mbox{arcsinh}\left(\tilde{\eta}
\sin\frac{p}{2}\right)
\right)\sqrt{\frac{1+\tilde{\eta}^2\sin^2\frac{p}{2}}{\left(1+\tilde{\eta}^2\right)\sin^2\frac{p}{2}}}\eea
and ${}_2F_1(a,b;c;z)$ is the Gauss' hypergeometric function.

As one can see from (\ref{fcd}), the leading finite-size
correction to $\mathcal{C}_{\eta}^{d,j}$ is exponentially small
for $J_1>>g$.

Now, let us consider two particular cases as examples.

For $j=1$, (\ref{fcd}) simplifies to \bea\label{fcd1}
&&\mathcal{C}_{\eta}^{d,1}\approx
\frac{3c_\Delta^{d,1}\pi\sqrt{1+\tilde{\eta}^2}}{4\tilde{\eta}^2\left(\tilde{\eta}^2+\csc^2\frac{p}{2}\right)}
\Bigg\{(2+\tilde{\eta}^2(1-\cos p))\csc^2\frac{p}{2}\times
\\ \nn &&\Bigg[\mathbf{K}\left(\frac{1}{1+\frac{\csc^2\frac{p}{2}}{\tilde{\eta}^2}}\right)
-\mathbf{E}\left(\frac{1}{1+\frac{\csc^2\frac{p}{2}}{\tilde{\eta}^2}}\right)\Bigg]
\\ \nn &&+\Bigg[4\Bigg(2(2+\tilde{\eta}^2(1-\cos p))
\mathbf{K}\left(\frac{1}{1+\frac{\csc^2\frac{p}{2}}{\tilde{\eta}^2}}\right)
\\ \nn &&-4(4+5\tilde{\eta}^2+2\tilde{\eta}^4-\tilde{\eta}^2(3+2\tilde{\eta}^2)\cos p)
\mathbf{E}\left(\frac{1}{1+\frac{\csc^2\frac{p}{2}}{\tilde{\eta}^2}}\right)\Bigg)
\\ \nn &&+4\tilde{\eta}^2(1+\cos p)\mathbf{E}\left(\frac{1}{1+\frac{\csc^2\frac{p}{2}}{\tilde{\eta}^2}}\right)f
\Bigg]\exp(-f)\Bigg\} .\eea

For $j=2$, (\ref{fcd}) reduces to \bea\label{fcd2}
&&\mathcal{C}_{\eta}^{d,2}\approx
\frac{32c_\Delta^{d,2}(1+\tilde{\eta}^2)}{15\tilde{\eta}^3\left(\tilde{\eta}^2+\csc^2\frac{p}{2}\right)^{3/2}}
\Bigg\{(2+\tilde{\eta}^2(1-\cos
p))\csc^2\frac{p}{2}\times
\\ \nn &&\Bigg[\sqrt{\tilde{\eta}^2+\csc^2\frac{p}{2}}\mbox{arctanh}
\left(\frac{\tilde{\eta}}{\sqrt{\tilde{\eta}^2+\csc^2\frac{p}{2}}}\right)
-\tilde{\eta}\ {}_2F_1\left(-\frac{1}{2},1;\frac{3}{2};
\frac{1}{1+\frac{\csc^2\frac{p}{2}}{\tilde{\eta}^2}}\right)\Bigg]
\\ \nn &&+8\Bigg[\sqrt{\tilde{\eta}^2+\csc^2\frac{p}{2}}
(2+\tilde{\eta}^2(1-\cos p))\mbox{arctanh}
\left(\frac{\tilde{\eta}}{\sqrt{\tilde{\eta}^2+\csc^2\frac{p}{2}}}\right)
\\ \nn &&-\tilde{\eta}(2+2 \tilde{\eta}^2+\tilde{\eta}^4-\tilde{\eta}^2(2+\tilde{\eta}^2)
\cos p)\ {}_2F_1\left(-\frac{1}{2},1;\frac{3}{2};
\frac{1}{1+\frac{\csc^2\frac{p}{2}}{\tilde{\eta}^2}}\right)
\\ \nn &&+\tilde{\eta}^3(1+\cos p)\ {}_2F_1\left(-\frac{1}{2},1;\frac{3}{2};
\frac{1}{1+\frac{\csc^2\frac{p}{2}}{\tilde{\eta}^2}}\right)
f\Bigg]\exp(-f)\Bigg\}.\eea

\setcounter{equation}{0}
\section{Giant magnons on $\eta$-deformed $AdS_5\times S^5$ and primary scalar operators}

It was shown in \cite{B0215} that the normalized semiclassical
structure constants for the case at hand are given by
\bea\label{c3prsf}
\mathcal{C}_{\eta}^{pr,j}&=&\frac{2\pi^{\frac{3}{2}}c_{\Delta}^{pr,j}\Gamma
(\frac{j}{2}) (1-v^2 \kappa^2)^{\frac{j-1}{2}}} {\Gamma
(\frac{1+j}{2})\sqrt{\kappa^2(1+\tilde{\eta}^2 \kappa^2)}}
\Bigg\{\left[1-\frac{(1+jv^2)\kappa^{2}}{1+j}\right]\times
\\ \nn &&F_1 \left(\frac{1}{2},\frac{j}{2},\frac{1-j}{2};1;\frac{\tilde{\eta}^2(1-v^2)\kappa^2}
{1+\tilde{\eta}^2\kappa^2},\frac{(1+\tilde{\eta}^2)(1-v^2)\kappa^2}
{(1+\tilde{\eta}^2\kappa^2)(1-v^2\kappa^2)}\right)
\\ \nn &&-(1-v^2\kappa^2) F_1 \left(\frac{1}{2},\frac{2+j}{2},-\frac{1+j}{2};1;\frac{\tilde{\eta}^2(1-v^2)\kappa^2}
{1+\tilde{\eta}^2\kappa^2},\frac{(1+\tilde{\eta}^2)(1-v^2)\kappa^2}
{(1+\tilde{\eta}^2\kappa^2)(1-v^2\kappa^2)}\right)\Bigg\} ,\eea or
in our notations

\bea\label{c3prf} \mathcal{C}_{\eta}^{pr,j}&=&
\frac{2\pi^{\frac{3}{2}}c_{\Delta}^{pr,j}\Gamma (\frac{j}{2})
(1-v^2 W)^{\frac{j-1}{2}}} {\Gamma
(\frac{1+j}{2})\sqrt{W(1+\tilde{\eta}^2 W)}}
\Bigg\{\left[1-\frac{(1+jv^2)W}{1+j}\right]\times
\\ \nn &&F_1 \left(\frac{1}{2},\frac{j}{2},\frac{1-j}{2};1;\frac{\tilde{\eta}^2(1-v^2)W}
{1+\tilde{\eta}^2 W},1-\epsilon\right)
\\ \nn &&-(1-v^2 W) F_1 \left(\frac{1}{2},\frac{2+j}{2},-\frac{1+j}{2};1;\frac{\tilde{\eta}^2(1-v^2)W}
{1+\tilde{\eta}^2 W},1-\epsilon\right)\Bigg\}.\eea

Taking the limit $\epsilon\to 0$ in (\ref{c3prf}) and using
(\ref{vW})-(\ref{v0}), one finds that to the leading order
$\mathcal{C}_{\eta}^{pr,j}$ becomes
\bea\label{appj}
&&\mathcal{C}_{\eta}^{pr,j}\approx
\frac{c_{\Delta}^{pr,j}\pi\Gamma(\frac{j}{2})}{\Gamma
(\frac{1+j}{2})}
\Bigg\{\Bigg[\frac{2\pi^{1/2}\sqrt{1+\tilde{\eta}^2}}
{j^2\tilde{\eta}^2\left(2+\tilde{\eta}^2(1-\cos
p)\right)}\Bigg(\frac{1+\tilde{\eta}^2} {\tilde{\eta}^2+\csc^2
p/2}\Bigg)^{\frac{j-1}{2}}
\\ \nn &&
{}_2F_1\left(\frac{1}{2},-\frac{1+j}{2};1;1\right)
\Bigg(2(1+j){}_2F_1\left(\frac{3}{2},\frac{j}{2};\frac{1+j}{2};
\frac{\tilde{\eta}^2}{\tilde{\eta}^2+\csc^2 p/2}\right)
\\ \nn &&-(2+j(2+\tilde{\eta}^2)-j\tilde{\eta}^2\cos p)
{}_2F_1\left(\frac{1}{2},\frac{j}{2};\frac{1+j}{2};
\frac{\tilde{\eta}^2}{\tilde{\eta}^2+\csc^2 p/2}\right) \Bigg)
\\ \nn &&+\Bigg[\Bigg(\Gamma\left(\frac{j}{2}-1\right)
\left(\frac{1+\tilde{\eta}^2}{\tilde{\eta}^2+\csc^2
p/2}\right)^{\frac{1+j}{2}} (2+\tilde{\eta}^2(1-\cos
p))
\\ \nn &&\times
\Bigg(-(1+\tilde{\eta}^2)\Big(4+2j(1-j)+\tilde{\eta}^2
\Big(4+2j(1-2j)-\frac{(2-j)j \cot^2 p/2}{\tilde{\eta}^2+\csc^2
p/2}(1+f)\Big)\Big)
\\ \nn &&\times \ {}_2F_1\left(-\frac{1}{2},\frac{j}{2};\frac{1+j}{2};
\frac{\tilde{\eta}^2}{\tilde{\eta}^2+\csc^2 p/2}\right)+
\Bigg(4(1+\tilde{\eta}^2)+j\Big(2(1-j) \\ \nn
&&+\tilde{\eta}^2\Big(4(1-j)+\frac{(2-j)\cot^2
p/2}{\tilde{\eta}^2+\csc^2 p/2}(1-f)\Big)+
\\ \nn &&
\frac{\tilde{\eta}^4\cot^2 p/2\Big(6-4j+\frac{(j-2)\cot^2
p/2}{\tilde{\eta}^2+\csc^2 p/2}(1+f)\Big)}{\tilde{\eta}^2+\csc^2
p/2} \Big)\Bigg)
{}_2F_1\left(\frac{1}{2},\frac{j}{2};\frac{1+j}{2};
\frac{\tilde{\eta}^2}{\tilde{\eta}^2+\csc^2 p/2}\right)
\Bigg)\Bigg] \\ \nn
&&/\Bigg(\Gamma\left(\frac{3+j}{2}\right)\tilde{\eta}^2(1+\tilde{\eta}^2)^{3/2}
\Biggr)+\Bigg(8\Gamma\left(1+\frac{j}{2}\right)\cos^2 p/2
\Bigg(\frac{1+\tilde{\eta}^2} {\tilde{\eta}^2+\csc^2
p/2}\Bigg)^{\frac{j+1}{2}}
\\ \nn &&\times \Bigg(\frac{2\ {}_2F_1\left(\frac{1}{2},\frac{j}{2};\frac{1+j}{2};
\frac{\tilde{\eta}^2}{\tilde{\eta}^2+\csc^2 p/2}\right)
}{2+\tilde{\eta}^2(1-\cos
p)}-{}_2F_1\left(-\frac{1}{2},\frac{j}{2};\frac{1+j}{2};
\frac{\tilde{\eta}^2}{\tilde{\eta}^2+\csc^2
p/2}\right)\Bigg)\Bigg) \\
\nn
&&/\left(\Gamma\left(\frac{3+j}{2}\right)\sqrt{1+\tilde{\eta}^2}
\right)f \Bigg]\exp(-f) \Bigg\}.\eea

As in the previously considered case, the leading finite-size
correction to $\mathcal{C}_{\eta}^{pr,j}$ is exponentially small
for $J_1>>g$.

For the simplest case $j=1$, (\ref{appj}) reduces to
\bea\label{app1} &&\mathcal{C}_{\eta}^{pr,1}\approx 2\
c_{\Delta}^{pr,1}\pi\frac{\sqrt{1+\tilde{\eta}^2}}{\tilde{\eta}^2}
\Bigg\{2\
\mathbf{E}\left(\frac{\tilde{\eta}^2}{\tilde{\eta}^2+\csc^2
p/2}\right)-\frac{2+\tilde{\eta}^2\sin^2
p/2}{1+\tilde{\eta}^2\sin^2 p/2}
\\ \nn &&\times\mathbf{K}\left(\frac{\tilde{\eta}^2}{\tilde{\eta}^2+\csc^2
p/2}\right)+\csc^2 p/2 \Bigg[\Bigg(2(1+\tilde{\eta}^2\sin^2
p/2)(8+7\tilde{\eta}^2+2\tilde{\eta}^4
\\ \nn &&-\tilde{\eta}^2(5+2\tilde{\eta}^2) \cos p)
\mathbf{E}\left(\frac{\tilde{\eta}^2}{\tilde{\eta}^2+\csc^2
p/2}\right)-(16+18 \tilde{\eta}^2+7\tilde{\eta}^4
\\ \nn &&-2\tilde{\eta}^2(7+4\tilde{\eta}^2)\cos p+\tilde{\eta}^4\cos 2p\Bigg)
\mathbf{K}\left(\frac{\tilde{\eta}^2}{\tilde{\eta}^2+\csc^2
p/2}\right)/\left(\tilde{\eta}^2+\csc^2 p/2\right)^2
\\ \nn &&+\Bigg(8\tilde{\eta}^2\cos^2 p/2\Bigg(\mathbf{K}\left(\frac{\tilde{\eta}^2}{\tilde{\eta}^2+\csc^2
p/2}\right)-(1+\tilde{\eta}^2\sin^2
p/2)\mathbf{E}\left(\frac{\tilde{\eta}^2}{\tilde{\eta}^2+\csc^2
p/2}\right)\Bigg)
\\ \nn &&\times \sin^4
p/2\Bigg)/(1+\tilde{\eta}^2\sin^2 p/2)^2 f \Bigg]\exp(-f)\Bigg\}
.\eea

\setcounter{equation}{0}
\section{Concluding Remarks}
In this letter, in the framework of the semiclassical approach, we
computed the leading finite-sizes effects on the normalized
structure constants in some three-point correlation functions in
the $\eta$-deformed $AdS_5\times S^5$, expressed in terms of the
conserved string angular momentum $J_1$ and the worldsheet
momentum $p$. Namely, we found the leading finite-size corrections
on the structure constants in three-point correlators  of two
heavy giant magnons' string states and the following two light
states:
\begin{enumerate}
\item{Dilaton operator with nonzero momentum ($j\ge 1$)}
\item{Primary scalar operators.}
\end{enumerate}

It is interesting to see what happens when we take the limit
$\tilde{\eta}\to 0$ when the deformation disappears, and the dual
field theory is known - $\mathcal{N}=4$ SYM.

For the case of dilaton operator with nonzero momentum $j$ one
finds \bea\label{dun} &&\mathcal{C}^{d,j}\approx \frac{2
c_\Delta^j \pi \Gamma\left(2+j/2\right)\Gamma\left(j/2\right)(\sin
p/2)^{1+j}}{(1+j)
\Gamma\left(\frac{1+j}{2}\right)\Gamma\left(\frac{5+j}{2}\right)}\times
\\ \nn &&\left[j-(2(4+(1-j)j(1+\cos p))-2j(1+j)(1+\cos p)f)\exp(-f)\right],\eea
where now (\ref{f}) becomes \bea\label{fud} f=\frac{J_1}{g\sin
p/2}+2=\frac{2\pi J_1}{\sqrt{\lambda}\sin p/2}+2,\eea and
$\lambda$ is the  t'Hooft coupling constant (the string tension
$T$ and the  t'Hooft coupling in $\mathcal{N}=4$ SYM are related
to each other as $T=\frac{\sqrt{\lambda}}{2\pi}$).

For the case of primary scalar operators we derive \bea\label{upr}
&\mathcal{C}^{pr,j}\approx 16 \pi c_{\Delta}^{pr,j}
\frac{\Gamma\left(j/2-1\right)\Gamma\left(j/2\right)}{\Gamma\left(\frac{j-1}{2}\right)
\Gamma\left(\frac{j+3}{2}\right)}
\sin^{j+1}\left(\frac{p}{2}\right)\exp(-f),\eea where $f$ is the
same as in (\ref{fud}).

A possible issue to solve is to try to extend the results obtained
here to the case of dyonic giant magnon states with two nonzero
conserved angular momenta $J_1$ and $J_2$. We hope to report on
this in the near future.

\section*{Acknowledgements}
This work is partially supported by the NSFB grant DFNI T02/6.

\end{document}